\documentclass[preprintnumbers,amsmath,amssymb]{revtex4}

\usepackage{amsmath}    
\usepackage{graphicx}   
\usepackage{color}

\pdfoutput=1
\begin{document}

\title{Quantum dice}
\author{Massimiliano Sassoli de Bianchi}
\affiliation{Laboratorio di Autoricerca di Base, 6914 Carona, Switzerland}\date{\today}
\email{autoricerca@gmail.com}   

\begin{abstract}

In a letter to Born, Einstein wrote~\cite{Letter}: ``Quantum mechanics is certainly imposing. But an inner voice tells me that it is not yet the real thing. The theory says a lot, but does not really bring us any closer to the secret of the `old one.' I, at any rate, am convinced that \emph{He} does not throw dice.'' In this paper we take seriously Einstein's famous metaphor, and show that we can gain considerable insight into quantum mechanics by doing something as simple as rolling dice. More precisely, we show how to perform measurements on a single die, to create typical quantum interference effects, and how to connect (entangle) two identical dice, to maximally violate Bell's inequality.\\


\end{abstract}

\maketitle

\section{Introduction}
\label{intro}

Quantum mechanics is a probabilistic theory, as most of its predictions are irreducibly statistical. It is therefore understandable that the first attempts to clarify its content made use of the well-tested concept of \emph{statistical ensembles}, describing identical abstract copies of the system under consideration, each of which would represent a different state in which the system might be found to be in. This statistical \emph{ensemble interpretation} of quantum physics was originally held by Albert Einstein, and subsequently supported by a number of authors, like for instance Leslie E. Ballentine~\cite{Ball}. 

We can summarize the core of this view by directly quoting Einstein~\cite{Eins}: ``The attempt to conceive the quantum-theoretical description as the complete description of the individual systems leads to unnatural theoretical interpretations, which become immediately unnecessary if one accepts the interpretation that the description refers to ensembles of systems and not to individual systems.''

In other terms, according to the statistical ensemble interpretaton, a state vector $|\psi\rangle$ doesn't describe the state of an individual system, but a more abstract entity: an ensemble of identical copies of the same system, each of which is in a different possible state. This would mean for instance that when we write $|\psi\rangle$ in the form of a superposition:
\begin{equation}
\label{superposition}
|\psi\rangle = \sum_{i=1}^{n} \alpha_i |a_i\rangle, \quad \sum_{i=1}^n |\alpha_i|^2=1, 
\end{equation}

\noindent where the $|a_i\rangle$, $i=1,\dots ,n$, are the eigenstates of a given observable $A$ (a self-adjoint operator), one should not understand it as if it was a real, actual state, describing a condition in which the individual system would nonsensically be, at once, in all these different mutually exclusive states, but simply as a convenient mathematical notation expressing the fact that, following a long series of measurements of the observable $A$, on identically prepared systems (the preparation being described by the vector $|\psi\rangle$), the systems will be found to be in one of the eigenstates $|a_i\rangle$,  $|a_i|^2 100 \%$ of the time. 

Of course, from a purely instrumentalistic point of view, there are no problems in adopting the minimalistic view that quantum theory is not about individual systems, but about statistical ensembles of similarly prepared systems. Indeed, it is a matter of fact that when experimenters measure a physical quantity in the laboratory, on a quantum system, what they do is precisely to repeat the same experiment a large number of times, on identically prepared systems, in order to calculate probabilities as limits of relative frequencies of outcomes. 

In other terms, there certainly exists at least one uncontroversial ensemble to which the state vector $|\psi\rangle$ refers to: the ensemble of identically prepared quantum entities which are subject to a series of identical measurements, as well as the ensemble of outcomes associated with them. Problems however begin when one quits a purely instrumentalistic-empiricistic view and begins to wonder what could be the reality of a microscopic quantum entity, like for instance an electron. 

If we consider the paradigmatic example of classical statistical mechanics, we can observe that the statistical ensemble this theory deals with is a \emph{purely theoretical construct}, resulting from the existence of another kind of ensemble, which instead is very concrete: the ensemble of identical microscopic entities forming the macroscopic system under consideration (for example the identical molecules forming a classical ideal gas). Each of these microscopic subsystems possesses a well-defined state, and each possible combination of the subsystems' individual states defines a specific state of the macroscopic system. But since we have no access, in terms of knowledge, to the state of each individual microscopic subsystem, we cannot have access to the actual state of the macroscopic system, which therefore can only be described in probabilistic terms, by means of an abstract statistical ensemble.

If quantum mechanics was just a theory dealing with systems formed by sub-entities possessing well-defined states, we could assume that the statistical content of $|\psi\rangle$ could be traced back, in a way or another, to our lack of knowledge about the different individual states in which the system's components are in, in a sort of generalization of a state of statistical mechanics. But such a view is difficult (if not impossible) to maintain if we consider our today's ability  of performing experiments also with a single microscopic entity at a time -- like for examples single neutrons in Rauch's celebrated interferometry experiments~\cite{Rauch} -- and when we do so, we still necessitate to use a pure probabilistic language to conveniently describe the outcomes of the measurements.

So, if we take seriously our ability to perform measurements on individual entities of a non-composite kind, it seems we are forced to conclude that the only possible origin of the statistical ensemble associated with $|\psi\rangle$ is in the series of experiments we perform on identically prepared systems. But if experiments are performed on identically prepared systems, i.e., on identical systems which are all exactly in the same condition, how is it possible that each single experiment can exhibit a potentially different outcome?

Consider for instance a marble moving rectilinearly on a table, characterized by a spatial position $x_0$ and velocity $v_0$ of its center of mass, at time $t=0$, and assume that we perform the experiment consisting in observing its position at a subsequent time $t_1$, and that the outcome of the experiment is $x_1$. If we consider an ensemble of identically prepared systems, i.e., of identical marbles all prepared in exactly the same state $(x_0,v_0)$, at time $t=0$, and perform on each of them the same position measurement at time $t=1$, the outcome will always be $x_1$. In other words, even though we have an ensemble of experiments, we only have a single outcome! 

Different from the quantum case, the state of a marble, when we consider it in relation to a position measurement, cannot be described as a superposition of different possible outcome. And this means that an ensemble of experiments performed on identically prepared systems is a necessary but certainly not a sufficient condition to obtain a statistical description of the system under consideration. 

The above is of course well known, and hidden variables theories have been attempted precisely in the hope of making up for this inconvenience. But associating hidden variables to, say, an electron's state, is about hypothesizing that the electron would be a sort of classical composite entity, with the hidden variables expressing our ignorance regarding the \emph{actual} states in which its different subcomponents would be in. In other terms, the typical ``hidden variables hypothesis'' is that when we prepare the system in a state $|\psi\rangle$, we have no practical control on the actual values taken by these hidden variables, which are responsible for the final outcome of the experiment. 

This idea that we need additional variables to describe the state of a quantum entity, in addition to the specification of $|\psi\rangle$, is of course very natural, and if proven correct it would provide a complete solution to the measurement problem, much in the spirit of a classical statistical theory. But such idea has encountered the immovable obstacle of the celebrated No-Go theorems~\cite{Gleason, Kochen, Bell}. Also, a hidden variable theory, with the hidden variables referring to our ignorance about the actual condition of the system, should  be described by a probabilistic theory obeying the classical Kolmogorovian axioms, whereas these axioms are disobeyed by quantum mechanics.~\cite{Accardi, Pitowski}

Clearly, all these problems turn around the fundamental question of giving a sensible meaning to a notion of probability associated with individual physical systems the state of which is assumed to be completely known. In other terms, the fundamental question we must ask is: can we understand probabilities not as quantifiers of our ignorance about the state of the system, but as quantifiers of our ignorance about something else? What would it be then this ``something else''? And, would it characterize in some objective way some of the features of the system under consideration? 

It is the purpose of the present article to provide a simple and clarifying answer to this fundamental question, on the basis of Aerts' \emph{hidden-measurement approach},~\cite{Aerts4, Aerts4b, Aerts7, Aerts10} by analyzing an extremely simple and well-known physical system: a six-faces die. 

More precisely, in Sections~\ref{Rolling a die is a quantum process}, \ref{A die with a Hilbert space representation} and \ref{Producing interferences with a  single die}, we show how to perform simple rolling experiments on a single die and describe the outcomes by means of the Born rule and the projection postulate, and how a single die can actually interfere with itself and violate the classical law of total probabilities. Furthermore, in Sec.~\ref{Violating Bell's inequality with two entangled dice}, we show how to connect (entangle) two identical dice and perform coincidence experiments that are capable to maximally violating Bell's inequality. Finally, in Sec.~\ref{Concluding remarks}, we offer some concluding remarks.

\section{Rolling a die is a quantum process}
\label{Rolling a die is a quantum process}

Undoubtedly, one of the most typical examples of a probabilistic experiment is the rolling of a die. If we ask what is the probability ${\cal P}(i)$ of obtaining the number $i\in \{1,2,3,4,5,6\}$, we can answer by simply applying Laplace's classical definition of probabilities: the ratio of favorable cases to all possible cases. Considering that a standard die has six different faces, if it is a fair die this ratio is simply: ${\cal P}(i) =1/6$, $\forall i\in \{1,2,3,4,5,6\}$.

A fundamental point in Laplace's definition of probability is the assumption that none of these possible cases has to be favored, in whatsoever way, by the selection procedure. And this means that, in the case of the die, the rolling process has to be genuinely \emph{random}, in the sense that it has to be such that it cannot distinguish between the different faces of the die (Jaynes' principle of indifference).

An important question to be asked is the following: What is the fundamental difference between the probabilities delivered by the experiment consisting in rolling a die and those delivered by a typical quantum measurement? Surprisingly, as we shall see, there are no fundamental differences, and the example of the die, if carefully analyzed, is actually able to provide all the important answers regarding a plausible origin of probabilities in quantum mechanics.

Let us start by observing that a die, when considered from the viewpoint of a rolling experiment, is a single, non-composite entity, about which we know in principle everything we need to know. In other terms, we are not here in a situation such that we could attach hidden variables to the die's state, so that if these variables were known they would allow us to dispense with the probabilities. Indeed, before rolling the die, if we really want we can perfectly well determine its exact state, for instance by taking a look to its upper face with respect to our hand's palm, and the exact location it occupies on the latter, at a given moment, but this knowledge, however complete it may be, is not going to help us in predicting the final outcome of the rolling experiment, when the die is thrown on the table. 

This is so because the main source of randomness in the experiment is not in our lack of knowledge about the initial state of the die, which we can assume to be fully known, but about the specific interaction taking place between the die and our hand which throws it, as well as, consequently, between the die and the table on which it will roll before exhibiting its final upper face. In other terms, if we really want to talk about hidden variables here, these will have to be attributed to the rolling experiment per se, and not to the initial die's state.

Of course, one can object that there is no fundamental difference between a rolling experiment with a die and our previous description of the marble moving rectilinearly on the table. Indeed, also the die, like the marble, follows a deterministic trajectory, which is just more complicated. So, the only important difference lies here in the fact that the actual die's trajectory depends on the unpredictable interaction with our hand, and this is the reason why we cannot easily predict the final outcome of the rolling experiment, whereas we can easily predict the positions of the marble on the table at whatever instant. 

In other terms, each rolling experiment with the die is a different experiment, in the sense that it expresses a different interaction between the die and the hand, and consequently between the die and the surface of the table, and this is why the result is in practice unpredictable. Each throw is per se a deterministic process, but since we lack knowledge about how the die is each time thrown, we can only describe the outcomes in probabilistic terms. 

A way to cope with this problem of indeterminacy is of course to construct an extremely precise machine able to throw the die always exactly in the same way, and for instance calibrate the machine (by varying for instance a certain parameter $\lambda$ which would control, say, the angular velocity and vertical speed with which the die is thrown) in such a way that if the die is placed on it with face $j$ up, it will also end its run on the table with the same face $j$ up. In other terms, by means of a very precise instrument, in replacement of our imprecise hand, we can produce a perfectly \emph{controlled} rolling experiment, and this time the probability of obtaining the number $i$ will be given by ${\cal P}_\lambda (i) = \delta_{ij}$, which is just another way to say that the outcome is now perfectly predetermined and we don't need any more to describe the experiment in probabilistic terms. 

Considering the above reasoning, it would appear that our statement that the rolling of a die can explain the nature of quantum probabilities, as there would be no essential differences between a rolling experiment and a quantum measurement, has been denied. But this is just because we are not considering the possibility of experimenting with a die from the right perspective. 

The right perspective we have here in mind is the one usually adopted in a typical gambling with dice at the casino, for instance in the game known as craps. Indeed, as is well known, in this kind of game one cannot roll the dice by using a machine, but one has instead to always use one's hand. The reason for this is precisely that the casino wants to prevent the player from taking a full control over the rolling experiment, as a full control would mean of course full predictability of the final outcome, and therefore a sure win. 

On the other hand, the casino doesn't forbid the player to know the initial states of the dice before the throw. This is so because a throw made by hand is not controllable by the player, and therefore knowing their initial states is not helpful in determining their final upper faces.

So, if we consider the roll of a die as an experiment of the  measurement kind, with the die being the physical entity which is measured, we can observe that from the viewpoint of the casino certain measurements are allowed, whereas certain others are strictly forbidden: we cannot perform a measurement with a high-precision machine, which is able to perfectly control the way the die is thrown (by fixing the parameter $\lambda$), but we can perform it by using a low-precision ``hand machine,'' which doesn't allow us to precisely control the way the die is actually thrown.   

Of course, the fact that certain die's experiments are forbidden from the casino's point of view doesn't mean they cannot be carried out: the point here is that a casino is only interested in having purely probabilistic outcomes, and that's why it \emph{imposes} to its players to carry out their rolling experiments by hand, and not by means of a high-precision instrument. Our point is that nature, like the casino, imposes a similar restriction when we deal with microscopic entities: we can only carry out experiments of the ``hand'' kind, and not of the ``high precision machine'' kind.

Having said that, and before analyzing some further the rolling experiment with a die, we need to make clear in which sense it can be considered a measurement. The question is: What is actually measured? The question is relevant because, as we said, we are here assuming that we perfectly know the state of the die before rolling it on the table: we know its mass, volume, its specific geometry, the total number of its faces, the material it is made of, its exact position and orientation on the palm of the hand, etc. So, what are we actually measuring here? 

Here again, we must adopt the viewpoint of the casino, for instance in a typical craps dice game. What matters, in the logic of the game are the upper faces obtained following a very specific rolling procedure which literally \emph{creates} a couple of upper faces (one for each die), the value of which will then determine the possible win or loss of the player. 

An important and subtle point here is about properly distinguishing a die's \emph{faces} from a die's \emph{upper face}. The six different faces of a die are of course always actually and stably existing, for as long as the die is not destroyed. In certain circumstances however, one of its six faces can temporarily become a so-called ``upper face.'' This happens each time a die is located on the flat surface of a game table. In that circumstance, the face having the highest gravitational energy corresponds to what is conventionally called the upper face of the die. 

Of course, a die can find itself located on the surface of a game table for a number of different reasons, one of which is surely the one of having taken part of a rolling experiment. Now, before such experiment is executed, each of the six \emph{actual faces} of the die are only \emph{potential upper faces}, as is clear that only one of the six faces will have the property of being ultimately placed upward, perpendicularly to the gravitational field. 

In other terms, in general, a rolling experiment, if properly understood, involves a pure \emph{creation aspect}. What is created are not the six faces of the die, which were existing also before the experiment and will continue to exist after it, but a specific upper face, which wasn't necessarily existing prior to the experiment (depending where the die was located). 

So, if we consider the rolling experiment with a die from the perspective of a process of creation of an upper face or, better, from the perspective of the measurement of the specific value written on the die's upper face, following a die's roll, we can observe that, similarly to the case of a quantum measurement on a microscopic system, the very process of measurement (observation) creates the property which is measured, i.e., it is the very measurement that actualizes the property which, prior to the measurement, was only existing in potential terms. 

What is important to note is that the presence of a process of \emph{actualization of potential} is a typical signature of quantum (or quantum-like) systems, exhibiting non-classical properties which can produce interference effects. And this means that, surprisingly, much of the ``weirdness'' of quantum physics is in fact already contained in the analysis of the most traditional examples that have been used to illustrate the classical probability calculus since the birth of probability theory, if we only interpret these examples as physical experiments testing specific properties, or measuring specific observables. 

To make this point fully explicit, in the next section we shall define specific rolling experiments on a very particular type of die, and show that the entire experimental situation can be easily described by means of a (real) Hilbert space structure, giving rise to typical quantum mechanical interference effects and therefore to a violation of the classical law of total probabilities.

\section{A die with a Hilbert space representation}
\label{A die with a Hilbert space representation}

\begin{figure}[!ht]
\centering
\includegraphics[scale =.5]{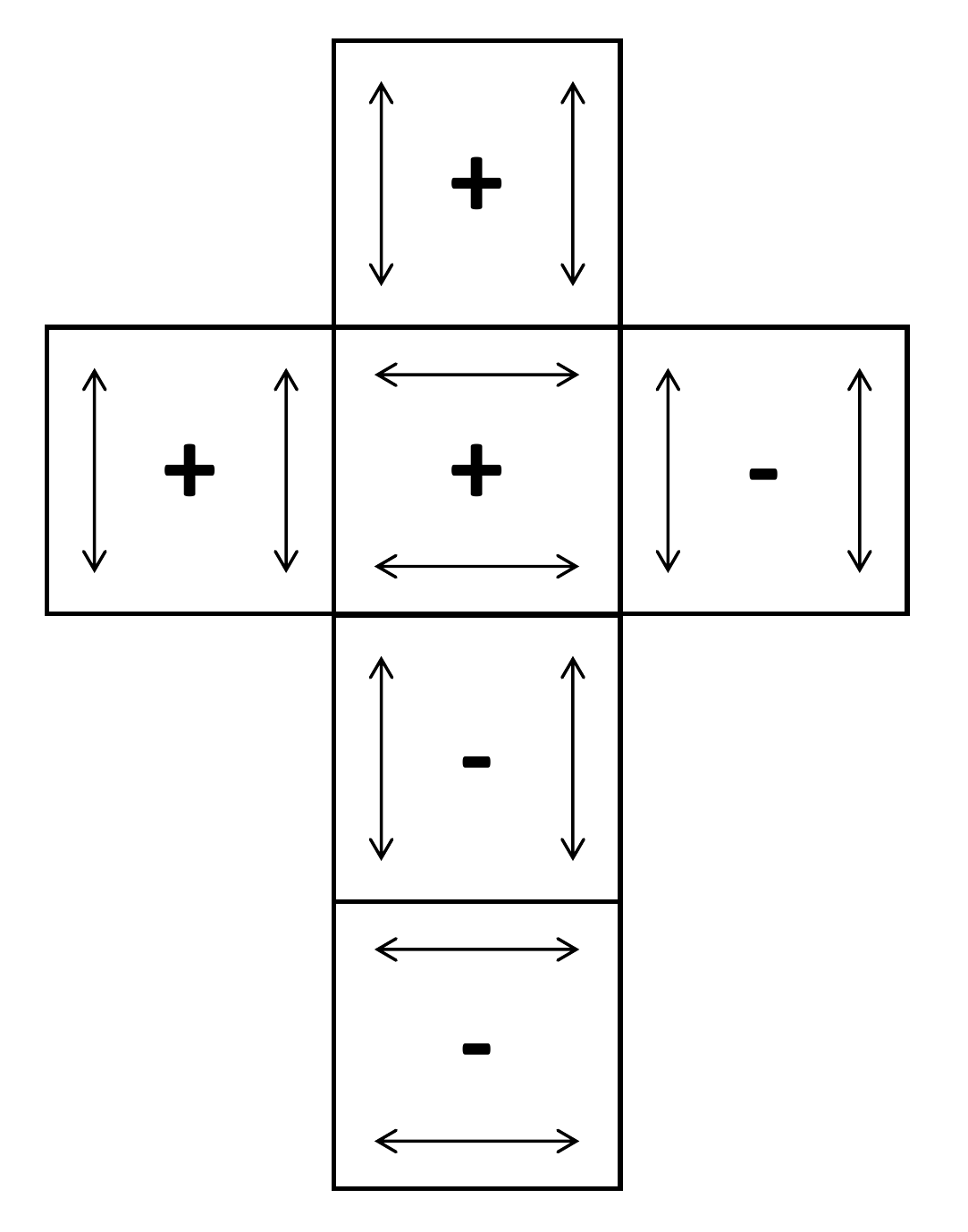}
\caption{The six-faces of the die, three of which show a ``$+$'' symbol, and the other three a ``$-$'' symbol. The surface of each face has a specific orientation, indicated by the two parallel left-right arrows, corresponding to the direction along which the die's face can easily glide on the flat surface of the game table.
\label{Quantum-die-faces}}
\end{figure}

The die we are going to consider is a traditional six-faces die. However, instead of numbering the faces, as usual, from $1$ to $6$, we shall only consider two numbers: $+1$ and $-1$, which for simplicity will be represented on the die's faces by the symbols ``$+$'' and ``$-$.'' As there is a total of six faces, the two symbols ``$+$'' and ``$-$'' will be repeated three times each on the die, in a way which is illustrated in Fig.~\ref{Quantum-die-faces}.

In addition to that, we shall consider that the surface of each face of the die is made of a particular material, which is able to slide with very low friction on the game table, along a specific direction, indicated on each die's face by two parallel left-right arrows (see Fig.~\ref{Quantum-die-faces}), but present a very high coefficient of friction as regards to the possibility of sliding in a direction perpendicular to that specified by the arrows. Note that the die is designed in such a way that the arrows of two opposite faces are always oriented in the same direction.

Apart from this peculiarity regarding the material with which the die's faces are made, the die is fair, in the sense that it is an object of perfectly homogeneous density. (To fix ideas, one can consider for instance that the surface of the game table is made of ice and that the two parallel arrows are two small metal blades, like those used on ice skates.)

The game table is a rectangular bi-dimensional surface, placed perpendicularly to the gravitational field, thus defining two orthogonal directions, corresponding to the two sides of the rectangle (which for simplicity will be considered of infinite length hereinafter), indicated as the $x$-direction and $z$-direction. (The reason of using the letter ``$z$'' instead of the letter ``$y$,'' as usual, will become clear later). 

\begin{figure}[!ht]
\centering
\includegraphics[scale =.6]{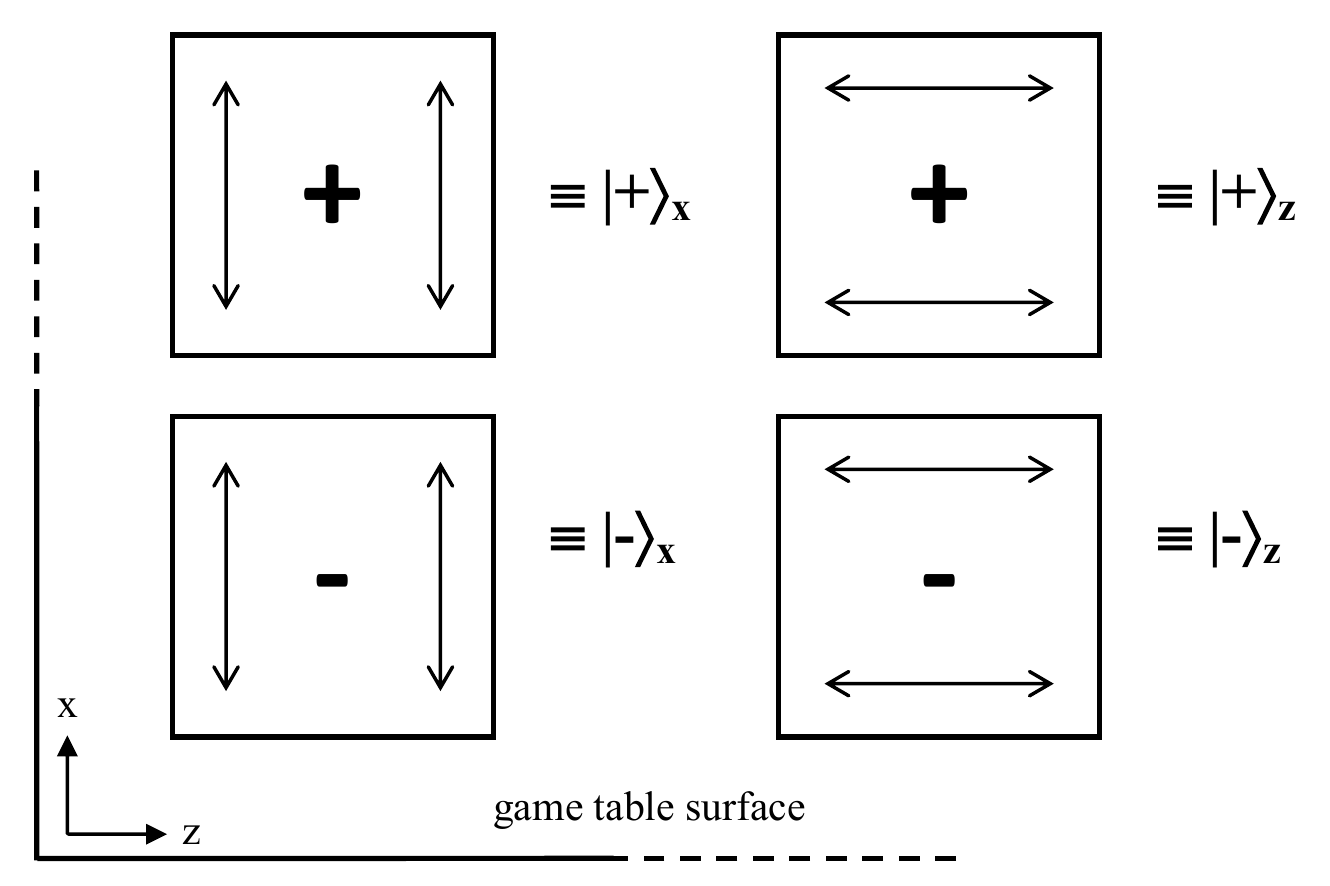}
\caption{The four different states of the die, corresponding to the two different possible orientations of the upper face's symbol with respect to the $x$ and $z$ directions, defined by the two sides of the game table. 
\label{die-states}}
\end{figure}

At the beginning of the game the die is placed on the surface of the game table with its upper face oriented either along the $z$-direction, or along the $x$-direction. Considering that only two different symbols are marked on the die's faces, this means that we only have to distinguish $4$ different states in which the die can be prepared: $|+\rangle_x$, $|-\rangle_x$, $|+\rangle_z$ and $|-\rangle_z$, as illustrated in Fig.~\ref{die-states}.

Let us observe that we have denoted these four different states by means of the typical quantum mechanical ket-notation, as an anticipation of the fact that we will be able to describe our measurements on the die-system by means of the Hilbert space formalism. This said, it is now time to define the observables that we are going to consider in relation to our die, and make precise how these observables are measured (i.e., observed) in concrete terms. 

\begin{figure}[!ht]
\centering
\includegraphics[scale =.4]{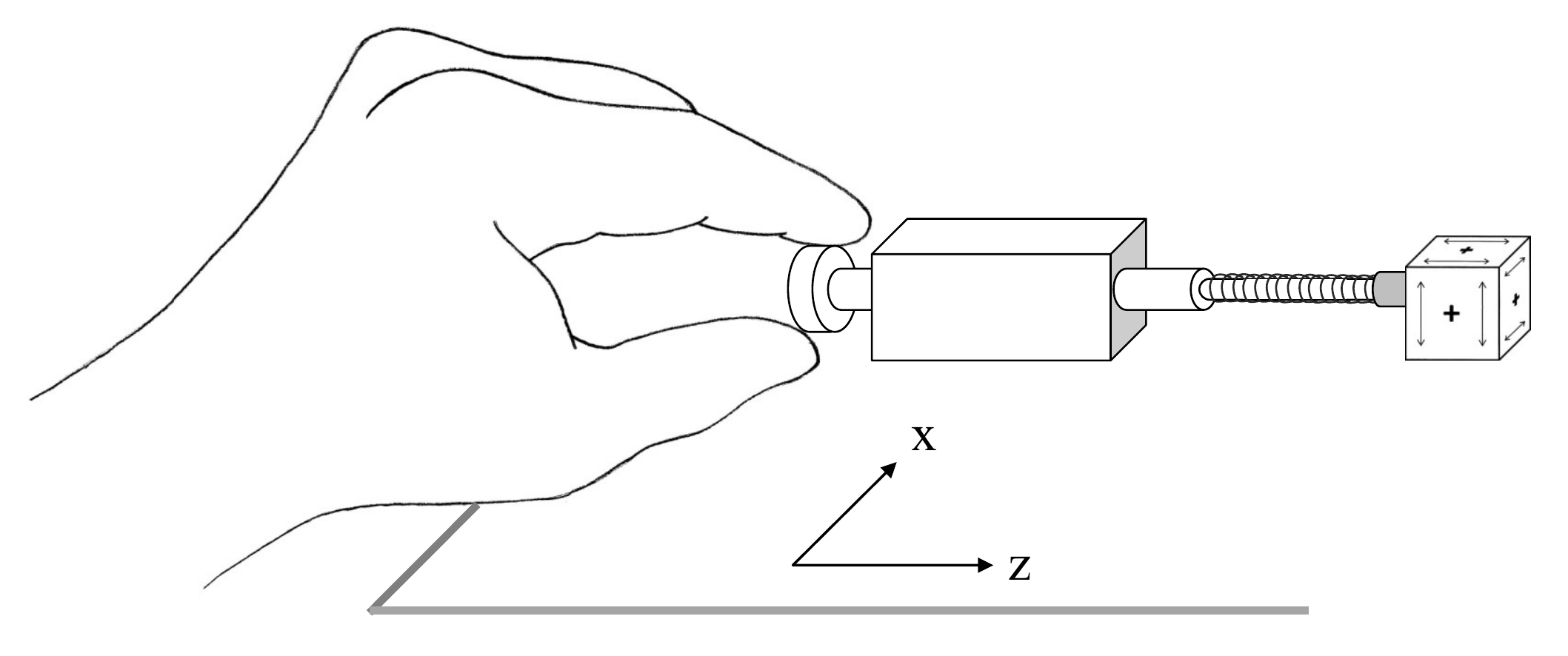}
\caption{The special shooter used by the experimenter allows the human operator to perfectly control the direction along which the die will be rolled, but not the impulsion which will be transferred to it.
\label{flipper-ball-shooter}}
\end{figure}

As usual in a game with dice, we are interested in observing the value exhibited by the die's upper face, as a result of a die's roll, i.e., as a result of a specific roll measurement. More precisely, we shall denote by $F_z$ the observable associated with the reading of the number marked on the die's upper face ($+1$, or $-1$), following a roll along the $z$-direction. The roll -- which in the following we shall simply call a $z$-\emph{roll} -- is performed by a human operator (the player, or the experimenter), by means of a special instrument, similar to a ``flipper ball shooter,'' thanks to which we ideally assume it is possible to produce a perfect roll of the die along the $z$-direction, as illustrated in Fig.~\ref{flipper-ball-shooter}.

More precisely, to measure $F_z$ the experimenter has to perfectly orient the shooter along the $z$-direction, placing it behind one of the two faces whose normal vector is parallel to the $z$-direction (which one of the two faces is actually chosen by the experimenter is irrelevant in terms of the outcome, because of the symmetry of the die), pull the knob in some arbitrary way (compressing in this way the spring in the mechanism), then release it, thus communicating a random (a priori unpredictable) impulsion to the die, which will after that either roll or not roll, according to its initial state. 

In fact, if the state of the die before the measurement of $F_z$ is either  $|+\rangle_z$ or $|-\rangle_z$, then, because of the low friction of the face in contact with the game table, the shooter will not be able to cause the die to roll, but only to glide on it, along the $z$-direction, for a certain time, until all translational kinetic energy will be converted into heat (we recall that the upper and lower faces of the die have the arrows oriented always in the same direction). In other terms, it is possible in this case to predict in advance, with certainty, the outcome of the measurement (without disturbing the system), which means that $|+\rangle_z$ and $|-\rangle_z$ are \emph{eigenstates} of $F_z$, with \emph{eigenvalues} $+1$ and $-1$, respectively. Using a standard Hilbertian representation, we can therefore write:
\begin{equation}
\label{eigenvalue equation-z}
F_z |\pm\rangle_z = \pm |\pm\rangle_z, \quad \phantom{.}_{z}\langle\pm | \mp\rangle_z = 0, \quad \phantom{.}_{z}\langle\pm | \pm\rangle_z =1.
\end{equation}

On the other hand, if the initial state of the die, before the measurement of $F_z$, is either $|+\rangle_x$ or $|-\rangle_x$, then, since the two (opposite) die's faces associated with these two states (see Fig.~\ref{die-states}) present an extremely high coefficient of friction with the game table, with respect to the z-direction, the die will not anymore slide following the action of the shooting machine, but roll along the $z$-direction (i.e., rotate around the $x$-direction). 

Of course, all initial rotational and translational energy will progressively be converted into thermal energy, so that in the end the die will stop and show a specific upper face. The dynamics of the rolling die can of course be very complex, but typically most of the energy communicated to it by the shooter will be initially transformed into rotational kinetic energy, then, because of the positive work performed by the friction forces, the rotational energy will be gradually transformed into translational kinetic energy and heat. 

For the purpose of our analysis, what is important to observe is that the die will roll for as long as the non-elastic effects associated with so-called \emph{rolling frictions} remain lower than the \emph{sliding frictions}, since in this case the die requires less energy to be moved by rolling than by sliding. But since two of the four faces involved in the rolling movement along the $z$-direction present an extremely low sliding friction, it is highly probable that the die will end its run sliding on one of them, before it will ultimately totally stops. 

In other terms, apart from exceptional circumstances, which we can simply ignore not to complicate our discussion unnecessarily, we can ideally assume that, following a $z$-roll, if the die's initial state is $|+\rangle_x$, or $|-\rangle_x$, then the final state will be either $|+\rangle_z$, or $|-\rangle_z$, i.e., an eigenstate of the $F_z$ observable.

Now, since the human operator has absolutely no control regarding the way the shooter will actually produce the roll of the die (apart from the rolling direction), it is clear that there is an equal probability of $1/2$ to obtain either the final state  $|+\rangle_z$, associated with the eigenvalue $+1$, or the final state $|-\rangle_z$, associated with the eigenvalue $-1$. This means that, with respect to the measurement of $F_z$, states $|+\rangle_x$ and $|-\rangle_x$ have to be considered as a \emph{superposition} of the two eigenstates $|+\rangle_z$ and $|-\rangle_z$ of $F_z$. A natural choice for their representation in terms of orthonormal states is therefore the following:
\begin{equation}
\label{superposition-x}
 |\pm\rangle_x = \frac{1}{\sqrt{2}}\left(|+\rangle_z \pm |+\rangle_z\right).
\end{equation}

So far, we have only considered the observable $F_z$, corresponding to an observation relative to the $z$-direction. In the same way, we can of course also consider the observable $F_x$, consisting in the observation of the number marked on the die's upper face following a roll along the $x$-direction ($x$-\emph{roll}), which is defined -- \emph{mutatis mutandis} -- likewise the $z$-roll, orienting in this case the shooter along the $x$-direction. Of course, the same discussion as per above can be repeated for $F_x$, and we can write: 
\begin{equation}
\label{eigenvalue equation-x}
F_x |\pm\rangle_x = \pm |\pm\rangle_x, \quad \phantom{.}_{x}\langle\pm | \mp\rangle_x = 0, \quad \phantom{.}_{x}\langle\pm | \pm\rangle_x =1.
\end{equation}
Again, with respect to the measurement of $F_x$, states $|+\rangle_z$ and $|-\rangle_z$ have to be considered as a \emph{superposition} of the two eigenstates $|+\rangle_x$ and $|-\rangle_x$ of $F_x$, so that we can write:
\begin{equation}
\label{superposition-z}
 |\pm\rangle_z = \frac{1}{\sqrt{2}}\left(|+\rangle_x \pm |+\rangle_x\right).
\end{equation}

Introducing the projection operators $P_{z,\pm}= |\pm\rangle_z \phantom{.}_{z}\langle\pm |$ onto the eigenspaces associated with states $|\pm\rangle_z$, and the projection operators $P_{x,\pm}= |\pm\rangle_x \phantom{.}_{x}\langle\pm |$, onto the eigenspaces associated with states $|\pm\rangle_x$, we can write: 
\begin{equation}
\label{observables}
 F_z = P_{z,+}- P_{z,-}, \quad F_x = P_{x,+}- P_{x,-}.
\end{equation}
Also, we can give a more explicit representation of these observables by setting:
\begin{equation}
\label{column-z}
|+\rangle_z = \left(\begin{array}{c} 1\\0\\ \end{array}\right), \quad |-\rangle_z = \left(\begin{array}{c} 0\\1\\ \end{array}\right).
\end{equation}
Then, according to (\ref{eigenvalue equation-z}), (\ref{superposition-x}) and (\ref{eigenvalue equation-x}), we have:
\begin{equation}
\label{column-x}
|+\rangle_x = \frac{1}{\sqrt{2}}\left(\begin{array}{c} 1\\1\\ \end{array}\right), \quad |-\rangle_x = \frac{1}{\sqrt{2}}\left(\begin{array}{c} 1\\-1\\ \end{array}\right) ,
\end{equation}
\begin{equation}
\label{Fz-and-Fx}
  F_z = \left(\begin{array}{cc}
                    1 & 0 \\
                    0 & -1
                  \end{array}\right), \quad F_x = \left(\begin{array}{cc}
                    0 & 1 \\
                    1 & 0
                  \end{array}\right),
\end{equation}
\begin{equation}
\label{Pz+-and-Pz-}
  P_{z,+}  = \left(\begin{array}{cc}
                    1 & 0 \\
                    0 & 0
                  \end{array}\right), \quad P_{z,-}  = \left(\begin{array}{cc}
                    0 & 0 \\
                    0 & 1
                  \end{array}\right),
\end{equation}
\begin{equation}
\label{Px+-and-Px-}
  P_{x,+}  = \frac{1}{2}\left(\begin{array}{cc}
                    1 & 1 \\
                    1 & 1
                  \end{array}\right), \quad P_{x,-}  = \frac{1}{2}\left(\begin{array}{rr}
                    1 & -1 \\
                    -1 & 1
                  \end{array}\right).
\end{equation}

Considering the way the two observables $F_z$ and $F_x$ have been operationally defined, in terms of the $z$-roll and $x$-roll experiments, respectively, it is easy to check that the above matrix representation, in association with the \emph{Born rule}, allows to consistently describe all the probabilities involved in the measurements of these two observables, which are the following: 
\begin{equation}
\label{probability1}
{\cal P} (|\pm\rangle_z\stackrel{z-\texttt{roll}}{\longrightarrow}|\pm\rangle_z)= \phantom{.}_{z}\langle\pm | P_{z,\pm} |\pm\rangle_z=|\!\phantom{.}_{z}\langle\pm |\pm\rangle_z|^2 = 1,
\nonumber
\end{equation}
\begin{equation}
\label{probability2}
{\cal P} (|\pm\rangle_z\stackrel{z-\texttt{roll}}{\longrightarrow}|\mp\rangle_z)= \phantom{.}_{z}\langle\pm | P_{z,\mp} |\pm\rangle_z=|\!\phantom{.}_{z}\langle\mp |\pm\rangle_z|^2 = 0,
\nonumber 
\end{equation}
\begin{equation}
\label{probability3}
{\cal P} (|\sigma\rangle_x\stackrel{z-\texttt{roll}}{\longrightarrow}|\rho\rangle_z)=
\phantom{.}_{x}\langle\sigma | P_{z,\rho} |\sigma\rangle_x
 =|\!\phantom{.}_{z}\langle\rho |\sigma\rangle_x|^2 =\frac{1}{2}, \quad\rho,\sigma\in\{+,-\}.
\end{equation}

Similarly, for the $x$-roll measurement, we have the probabilities: 
\begin{equation}
\label{probability1bis}
{\cal P} (|\pm\rangle_x\stackrel{x-\texttt{roll}}{\longrightarrow}|\pm\rangle_x)= \phantom{.}_{x}\langle\pm | P_{x,\pm} |\pm\rangle_x=|\!\phantom{.}_{x}\langle\pm |\pm\rangle_x|^2 = 1,
\nonumber
\end{equation}
\begin{equation}
\label{probability2bis}
{\cal P} (|\pm\rangle_x\stackrel{x-\texttt{roll}}{\longrightarrow}|\mp\rangle_x)= \phantom{.}_{x}\langle\pm | P_{x,\mp} |\pm\rangle_x=|\!\phantom{.}_{x}\langle\mp |\pm\rangle_x|^2 = 0,
\nonumber 
\end{equation}
\begin{equation}
\label{probability3bis}
{\cal P} (|\sigma\rangle_z\stackrel{x-\texttt{roll}}{\longrightarrow}|\rho\rangle_x)=
\phantom{.}_{z}\langle\sigma | P_{x,\rho} |\sigma\rangle_z
 =|\!\phantom{.}_{x}\langle\rho |\sigma\rangle_z|^2 =\frac{1}{2}, \quad\rho,\sigma\in\{+,-\}.
\end{equation}

Also, in accordance with the quantum mechanical \emph{projection postulate}, we can observe that the $z$-roll and $x$-roll experiments are to be considered \emph{ideal measurements}, as is clear that following the measurement of $F_z$ (resp. $F_x$), the initial state of the die-system is projected onto an eigenstate of $F_z$ (resp. $F_x$), a fact which can also be expressed by considering that the probabilities of finding the system either in state $|+\rangle_x$ or $|-\rangle_x$ (resp. $|+\rangle_z$ or $|-\rangle_z$), following a $z$-roll (resp. a $x$-roll), is equal to zero.

\section{Producing interferences with a single die}
\label{Producing interferences with a  single die}

The attentive reader will have certainly noticed that $F_z=\sigma_z$ and $F_x=\sigma_x$, where $\sigma_z$ and $\sigma_x$ are two of the three Pauli matrices (and this explains why we have unconventionally chosen letters $z$ and $x$ to denote the two rolling directions on the plane of the game table). Now, considering that (see any book of quantum mechanics) $[\sigma_x, \sigma_z]=-2i\sigma_y\neq 0$, it immediately follows that the two observables $F_z=\sigma_z$ and $F_x=\sigma_x$ are \emph{experimentally incompatible}, as the matrices representing them do not commute.

The existence of experimental incompatibility of certain observables is what distinguish, among other things, quantum physics from classical physics. More precisely, the presence of relations of non-commutation between certain observables is at the origin in quantum theory of so-called \emph{interference effects}, which in turn are responsible for a violation of the classical \emph{formula of total probability}. 

Let us show how interference effects, and consequently the violation of total probability, simply manifest in our measurements with the die. To do so, let us first generally observe that if $|\psi\rangle$ is the initial state of a given system, $A$ is a self-adjoint operator associated to a physical observable, and $P_{\alpha} $ is the projection operator associated with one of its eigenvalues $\alpha$, then, if $\alpha$ is the observed outcome of a measurement of $A$, according to the projection postulate the pre-measurement state $|\psi\rangle$ will ``collapse,'' following the measurement process, into the post-measurement state:
\begin{equation}
\label{projection}
|\psi_{\alpha}\rangle =  \frac{P_{\alpha}|\psi\rangle}{\sqrt{\langle\psi| P_{\alpha}|\psi\rangle}}. \end{equation}

Then, considering a second self-adjoint observable $B$, not necessarily commuting with $A$, we can ask what is the probability that the outcome of a measurement of $B$ would be one of its eigenvalues $\beta$, associated with the projection operator $P_{\beta} $, \emph{conditional} to the fact the previous measurement of $A$ produced $\alpha$ as an outcome. According to (\ref{projection}) and the Born rule, we know that such a conditional probability is given by:
\begin{equation}
\label{conditional-probability}
{\cal P}_{\psi}(B=\beta|A=\alpha) = \langle \psi_{\alpha} |P_{\beta} |\psi_{\alpha}\rangle =  \frac{\langle \psi | P_{\alpha}P_{\beta}P_{\alpha}|\psi\rangle}{\langle\psi| P_{\alpha}|\psi\rangle}.
\end{equation}

Considering then that ${\cal P}_{\psi}(A=\alpha)=\langle\psi| P_{\alpha}|\psi\rangle$ is the probability of obtaining the outcome $\alpha$ when $A$ is measured with the system in state $|\psi\rangle$, we can write:
\begin{equation}
\label{joint-probability}
{\cal P}_{\psi}(B=\beta|A=\alpha) {\cal P}_{\psi}(A=\alpha) = \langle \psi | P_{\alpha}P_{\beta}P_{\alpha}|\psi\rangle.
\end{equation}

This means that, by definition of a conditional probabilistic statement, the term on the right hand side of (\ref{joint-probability}) has to be interpreted as a \emph{joint probability} for the measurement of observables $A$ and $B$. However, since the two observables are not necessarily compatible, the joint probability is not here to be understood in the sense of the joint probability of two \emph{simultaneous} measurements, as $A$ and $B$ cannot in general be measured simultaneously, but as the joint probability of two \emph{sequential} measurements: 
\begin{equation}
\label{joint-probability-bis}
{\cal P}_{\psi}(A=\alpha \,\,\texttt{then}\,\, B=\beta) = \langle \psi | P_{\alpha}P_{\beta}P_{\alpha}|\psi\rangle.
\end{equation}

Defining the projection operator $P_{\bar \alpha}=\mathbb{I}-P_\alpha $, we can of course also write:
\begin{equation}
\label{joint-probability-tris}
{\cal P}_{\psi}(A\neq\alpha \,\,\texttt{then}\,\, B=\beta) = \langle \psi | P_{\bar \alpha}P_{\beta}P_{\bar \alpha}|\psi\rangle,
\end{equation}
and observing that:
\begin{equation}
\label{projection-relations}
P_\beta = \left(P_\alpha +P_{\bar \alpha }\right)P_\beta \left(P_\alpha +P_{\bar \alpha }\right)
 = P_\alpha P_\beta P_\alpha + P_{\bar \alpha}P_\beta P_{\bar \alpha}+P_\alpha P_\beta P_{\bar \alpha}+P_{\bar \alpha}P_\beta P_\alpha,
\end{equation}
we obtain from (\ref{joint-probability-bis}), (\ref{joint-probability-tris}) and (\ref{projection-relations}):

\begin{equation}
\label{quantum-total-probability}
{\cal P}_{\psi}(B=\beta) = {\cal P}_{\psi}(A=\alpha \,\,\texttt{then}\,\, B=\beta)
+ {\cal P}_{\psi}(A\neq\alpha \,\,\texttt{then}\,\, B=\beta)
+ 2 \Re \,\langle \psi |P_{\alpha}P_{\beta}P_{\bar \alpha}|\psi\rangle.
\end{equation}

Eq.~(\ref{quantum-total-probability}) can be considered as the quantum generalization of the classical formula of total probability. When observables $A$ and $B$ are compatible, that is, when they commute, then also the corresponding projection operators commute, and since $P_{\alpha}P_{\bar \alpha}=P_{\alpha }-P_{\alpha}^2 = 0$, the third term in (\ref{quantum-total-probability}), which is a typical interference term, vanishes (and we recover the classical formula of total probability).

Formula (\ref{quantum-total-probability}) being  general, it also applies in relation to the two non-commuting observables $F_z$ and $F_x$, associated with the observation of the upper face of our die, following a $z$-roll and a $x$-roll experiment, respectively. Therefore, exactly as for a quantum microscopic system, the die is able to produce interference effects, and consequently violate the law of total probability. 

Let us check this fact more explicitely, in a specific example. For this, we set $P_\beta =P_{z,+}$, $P_\alpha = P_{x,+}$, $P_{\bar \alpha} = P_{x,-}$, and $|\psi\rangle = |+\rangle_z$. Then, since a $z$-rolling experiment cannot change the upper face of the die, when its upper face is oriented along the $z$-direction (the die will only glide, instead of rolling, since also the face in contact with the game table is oriented in the gliding sense), we have: 
\begin{equation}
\label{example1}
{\cal P}_{|+\rangle_z}(F_z=+1)={\cal P} (|+\rangle_z\stackrel{z-\texttt{roll}}{\longrightarrow}|+\rangle_z) =1.
\end{equation} 

On the other hand, for the two joint-sequential probabilities, we have: 
\begin{eqnarray}
\label{example2}
{\cal P}_{|+\rangle_z}(F_x=+1\,\,\texttt{then}\,\, F_z=+1) &=& {\cal P} (|+\rangle_z\stackrel{x-\texttt{roll}}{\longrightarrow}|+\rangle_x)\,{\cal P} (|+\rangle_x\stackrel{z-\texttt{roll}}{\longrightarrow}|+\rangle_z)\nonumber\\
&=& \frac{1}{2}\cdot\frac{1}{2}=\frac{1}{4},
\end{eqnarray} 
\begin{eqnarray}
\label{example3}
{\cal P}_{|+\rangle_z}(F_x=-1\,\,\texttt{then}\,\, F_z=+1) &=& {\cal P} (|+\rangle_z\stackrel{x-\texttt{roll}}{\longrightarrow}|-\rangle_x)\,{\cal P} (|-\rangle_x\stackrel{z-\texttt{roll}}{\longrightarrow}|+\rangle_z)\nonumber\\
&=& \frac{1}{2}\cdot\frac{1}{2}=\frac{1}{4}.
\end{eqnarray} 

Now, as is clear that $1\neq \frac{1}{4} + \frac{1}{4}$, the die manifestly violates the classical total probability's formula, in accordance with the fact that the matrices associated with the $F_z$ and $F_x$ observables do not commute. On the other hand, according to (\ref{column-z}), (\ref{Pz+-and-Pz-}) and (\ref{Px+-and-Px-}), we obtain for the inteference term: 
\begin{equation}
\label{interference-term}
2 \Re \,\phantom{.}_{z}\langle + |P_{x,+}P_{z,+}P_{x,-}|+\rangle_z = \frac{1}{2}\begin{array}{cc}
                    (1 & 0) \\
                    \phantom{(1} & \phantom{0)}
                  \end{array}\left(\begin{array}{cc}
                    1 & 1 \\
                    1 & 1
                  \end{array}\right) \left(\begin{array}{cc}
                    1 & 0 \\
                    0 & 0
                  \end{array}\right) \left(\begin{array}{rr}
                    1 & -1 \\
                    -1 & 1
                  \end{array}\right) \left(\begin{array}{c} 1\\0\\ \end{array}\right)= \frac{1}{2}\nonumber,
\end{equation} 
which is precisely the value we need to add to (\ref{example2}) plus (\ref{example3}) in order to recover (\ref{example1}), in accordance with (\ref{quantum-total-probability}). 

We leave it to the reader to verify that (\ref{quantum-total-probability}) correctly describes the relation between marginal probabilities, joint-sequential probabilities and interference contributions, for other choices of the initial state and of the conditioning.

\subsection{Discussion}
\label{Discussion1}

For some readers it may come as a surprise that also macroscopic, ordinary systems, can behave in a quantum (or quantum-like) manner. This however is well known since a long time now, at least by foundational researchers, like those of the Geneva-Brussel school of quantum mechanics, whose approach to the foundation of physical theories originated from the pioneering work of Josef-Maria Jauch and Constantin Piron in Geneva,~\cite{Piron1, Piron2, Piron3}, and subsequently from that of Diederik Aerts and collaborators in Brussels~\cite{Aerts1, Aerts5, Aerts2, Aerts6, Aerts3, Aerts4b}.

It is important to observe that the quantum behavior of a macroscopic system, like a die, is not a consequence of its internal coherence, but of the way we have decided to actively experimenting with it, by means of some very specific experimental protocols. More precisely, the quantum behavior of a die, and of other macroscopic systems exhibiting a quantum (or quantum-like) structure, is a consequence of the fact that we are not conceiving our observations on the system (i.e., our measurements) only as processes of pure discovery, but also as processes of creation, i.e., processes through which we can create, in an unpredictable manner, the very quantities we are measuring (this is what is sometimes called the \emph{observer effect}. See~\cite{Sassoli-Observer} and the references cited therein). 

In our die-system this is has been done by considering the two observables $F_z$ and $F_x$, corresponding to the reading of the symbol marked on the upper face of the die, following particular rolling experiments. According to the ``rules of the quantum (and casino) game,'' these ``reading measurements'' cannot be performed by passively looking at the die on the table, but by means of procedures which require that the die is first rolled along the $z$ or $x$ directions (what we have called a $z$-roll and $x$-roll). These are, undoubtedly, readings of a very special kind! 

Of course, when experimenting with our die, we also have the possibility of looking at it in every moment, and directly ``see'' which one of the faces is in the upper position (if any). This possibility, of continuously monitoring the orientation of the die is in fact what confers the die-model its great explicative power, allowing for a full visualization of the measurement process, as it evolves (something we cannot do with microscopic entities).

However, such a possibility of continuously monitoring cannot be used here in practical terms, as an alternative to the $z$-roll and $x$-roll experiments, to determine the value of the quantum-like $F_z$ and $F_x$ ``upper-face'' observables, in the same way as when we play craps in a casino we must conveniently roll the dice for our possible win to be validated. 

Now, if we consider the die-system and its unusual ``roll measurements'' a meaningful structural analogy of what truly goes on, behind the scenes, during a quantum measurement with a microscopic entity, then we can highlight, as was done many years ago by Diederik Aerts~\cite{Aerts4, Aerts4b, Aerts7, Aerts10}, a very simple and physically transparent mechanism which would be at the origin of quantum probabilities. 

Indeed, coming back to our discussion in the Introduction, what our analysis of the die clearly shows is that we don't need an ensemble of entities to generate a statistical ensemble: a statistical ensemble can also be naturally attached to a given physical entity when, instead of considering its properties and states in purely static terms, we also understand them in \emph{dynamical terms}. 

What we don't have to forget is that although the state of an entity is a description of its actual properties, i.e., those properties whose actuality would be confirmed with certainty, should we decide to test them (i.e., to observe them), such a description also contains dynamical information about what the entity can possibly become if, when in a given state, we act on it in a certain way, according to a certain experimental protocol. 

If this protocol has a built-in random element, having its origin in the presence of some unavoidable fluctuations in the experimental context, then of course the becoming of the entity can only be described in probabilistic terms. Also, considering that when we act on a system we generally modify its state, it is natural to interpret the logical connectives subtending the probabilistic calculus of quantum systems not in a classical and static \emph{propositional} way, but in a strictly \emph{dynamical} way.~\cite{Smets}

This way of conceiving and interpreting the reality of quantum systems, in relation to the measurements we perform on them, is well illustrated in our experiments with the die. When for instance the die is in state $|+\rangle_z$, we can say that it possesses, in actual terms, the property of having what we may call a ``+'' $z$-\emph{upper face}. This we can say because we can predict with certainty (with probability equal to $1$) that the outcome of an observation of $F_z$ is $+1$, which is exactly what it is meant by possessing a ``+'' $z$-upper face.  

This certain prediction is possible because when the state is $|+\rangle_z$, and the system is operated through a $z$-roll, the fluctuations in the experimental context -- those produced by the random interaction of our hand with the shooter -- are not able to affect the final outcome. But this is not the case if we act on the system by means of a $x$-roll, i.e., if we measure $F_x$ instead of $F_z$. Indeed, in this case our lack of knowledge about the exact (deterministic) interaction which the hand selects, when pulling on the shooter, translates into our lack of knowledge about the final upper face exhibited by the die. 

In other terms, if, on the one hand, we can say that the die in state $|+\rangle_z$ possesses in \emph{actual} terms the ``+'' $z$-\emph{upper face} property, on the other hand we can only say that, when in such state, it possesses in \emph{potential} terms the ``+'' and ``$-$'' $x$-\emph{upper face} properties, and we can express this potentiality in more precise terms by writing the state of the system as the superposition: $|+\rangle_z = (|+\rangle_x + |-\rangle_x)/\sqrt{2}$.  

So, the state describes what the system \emph{is}, its actual properties, which correspond to those observations whose results we can predict with certainty, but also, indirectly, it describes what the system \emph{can possibly become}, when we observe properties which are not yet possessed by it, and therefore can only be created by the observational process, in a way which cannot be predicted in advance. 

This approach to the measurement problem has been called by Aerts the \emph{hidden-measurement approach}, where the term ``hidden'' refers to the deterministic interaction between the measured system and the measuring apparatus, which is selected in a random way during the measurement process, because of the presence of unavoidable and uncontrollable fluctuations in the experimental context.~\cite{Aerts4,Aerts4b,Aerts7,Aerts10}

More precisely, according to this approach (which has been substantiated in full mathematical terms by Aerts and Coecke~\cite{Aerts7,Coecke2}), to a given quantum measurement we can associate an entire collection of ``hidden'' deterministic measurements, and when the measurement is actually performed only one of these hidden-measurements does actually take place. Each one of these hidden deterministic measurements determine, in a unique way, a given outcome, but since we lack knowledge about which one is actually selected, we also lack knowledge about the final outcome. 

This is what would constitute the essential difference between classical probabilities, obeying Kolmogorov's axioms, and non-classical, quantum probabilities, disobeying Kolmogorov's axioms. The formers correspond to situations where the lack of knowledge is only about the state of the system, whereas the latter would correspond to situations of full knowledge of the system's state, but maximum lack of knowledge about the exact measurement interaction taking place between the system and the apparatus. 

What is interesting to observe is that between these two extremes, one can also describe intermediate pictures, giving rise to intermediate probabilities which can neither be fitted into a quantum probability model, nor into a classical probability model.~\cite{Accardi, Pitowski, Aerts3, Massimiliano2} 
 
A few additional comments are in order. In our description of the rolling experiment with the die, we have described measurements, probabilities and outcomes in terms of matrices and vectors in a real two-dimensional Hilbert space, using the projection postulate and the Born rule. However, to do so we have limited the possible states and observations on the system by only considering the two observables $F_z$ and $F_x$, relative to the two orthogonal directions $z$ and $x$, and we have also assumed that the system can only be prepared in one of the four states $|+\rangle_x$, $|-\rangle_x$, $|+\rangle_z$ and $|-\rangle_z$, as illustrated in Fig.~\ref{die-states}. In other terms, we have only considered a subset of all possible states of the die on the game table; a subset which is closed under the action of the observables $F_z$ and $F_x$. 

Of course, nothing prevents us from also defining more general observables $F_u$, associated to other directions $u$, different from $z$ or $x$. However, if we measure $F_u$ (by orienting the shooter along the $u$-direction) when the die-system is, say, in state $|+\rangle_z$, then if $u$ is not parallel or orthogonal to $z$, the post-measurement state will not in general be an eigenstate of $F_u$. This means that in most cases $F_u$ has to be considered a \emph{generalized observable}, and that the rolling experiments on the die-system cannot be generally described in terms only of so-called (von Neumann) ideal measurements and the projection postulate. 

More of course should be said about these important ideas, and the mathematical developments they have originated, particularly the difference between classical and quantum properties, classical and quantum probabilities, phase space and Hilbert space structures, as well as the intermediate structures corresponding to situations of partial (non-maximal) absence of knowledge, but this would go beyond the mostly didactical scope of the present paper. 

What we shall do instead, in the next section, is to show how it is possible to connect two identical dice, in order to create a double-die system which, like microscopic entangled systems, is able to violate Bell's inequality. In other terms, by ``playing'' with dice we can shed light not only on the origin of quantum interference effects and quantum probabilities, but also on the phenomenon of entanglement.

\section{Violating Bell's inequality with two entangled dice}
\label{Violating Bell's inequality with two entangled dice}

Before describing our system of two entangled dice, and show how we can perform experiments that will produce a violation of Bell's inequality, let us briefly recall what the latter is all about~\cite{Bell0, Clauser, Bell1, Merm} (see for instance~\cite{Vald} for a simple but general proof). Bell was able to write a mathematical inequality incorporating certain general assumptions about physical systems, so that if the inequality is found to be experimentally violated, then at least one of the assumptions used in its derivation must be wrong. Let us simply recall the expression of Bell's inequality, without proving it (we consider here the so-called CHSH generalization of it). 

On a given physical entity we assume that four different experiments can be performed: $e^A_a$, $e^A_{a'}$, $e^B_b$ and $e^B_{b'}$. Let us call $o^A_a$, $o^A_{a'}$, $o^B_b$ and $o^B_{b'}$ the outcomes associated to these experiments, which we assume can only take the two values $+1$ or $-1$. We also assume that experiments $e^A_a$ and $e^A_{a'}$ can be performed together with either of experiments $e^B_b$ and $e^B_{b'}$, thus defining additional \emph{coincidence} experiments: $e_{ab}^{AB}$,  $e_{ab'}^{AB}$, $e_{a'b}^{AB}$ and $e_{a'b'}^{AB}$. To each coincidence experiment $e_{cd}^{AB}$, $c\in\{a,a'\}$, $d\in\{b,b'\}$, one can associate the expectation value $E^{AB}_{cd}$ of the product of outcomes $o^A_c o^B_d$, by:
\begin{eqnarray}
\label{expectation value}
E^{AB}_{cd}&=&\sum {\cal P}_{cd}^{AB}(o^A_c,o^B_d) o^A_co^B_d\nonumber\\
&=& +{\cal P}_{cd}^{AB}(+1,+1) + {\cal P}_{cd}^{AB}(-1,-1) - {\cal P}_{cd}^{AB}(+1,-1) - {\cal P}_{cd}^{AB}(-1,+1),
\end{eqnarray}
where ${\cal P}_{cd}^{AB}(o^A_c,o^B_d)$ is the probability that the coincidence experiment $e_{cd}^{AB}$ yields the outcomes $(o^A_c,o^A_d)$. 

Assuming, as Bell did, that the experiments' outcomes are independently determined by some hidden variables, so that the expectation (\ref{expectation value}) can be written as the integral of the product of the two outcomes over these hidden variables (an hypothesis often referred to as \emph{Bell locality}), it is possible to prove the following relation~\cite{Bell1, Bell0}:
\begin{equation}
\label{Bell inequalities}
I\equiv |E^{AB}_{ab} - E^{AB}_{ab'}| + |E^{AB}_{a'b'} + E^{AB}_{a'b}|\leq 2.
\end{equation}

As is well known, (\ref{Bell inequalities}) is violated by certain quantum systems, like for instance those formed by two entangled spin-$1/2$ entities in a \emph{singlet (zero) spin state}, for which one can show that $I= 2\sqrt{2}>2$.~\cite{Aspect1, Aspect2, Massimiliano-elastic} In other terms, quantum systems formed by two entangled subsystems usually violate Bell's locality assumption, and this remains true even though the two subsystems are separated by a very large spatial distance. This means that no local physical theory, in the sense specified by Bell, can agree with all statistical implications of quantum mechanics, and that spatial separation doesn't imply \emph{experimental separation}. 

In order to gain some insight into the content of Bell's inequality, and understand what could be the reason of its violation by microscopic systems, like singlet spin states, we want now to show how two dice can be \emph{connected} to create a macroscopic entangled system which also violates (\ref{Bell inequalities}). This will shed some light into the nature of \emph{quantum correlations} (the ``spooky actions at a distance,'' as Einstein used to call them). 

\begin{figure}[!ht]
\centering
\includegraphics[scale =.5]{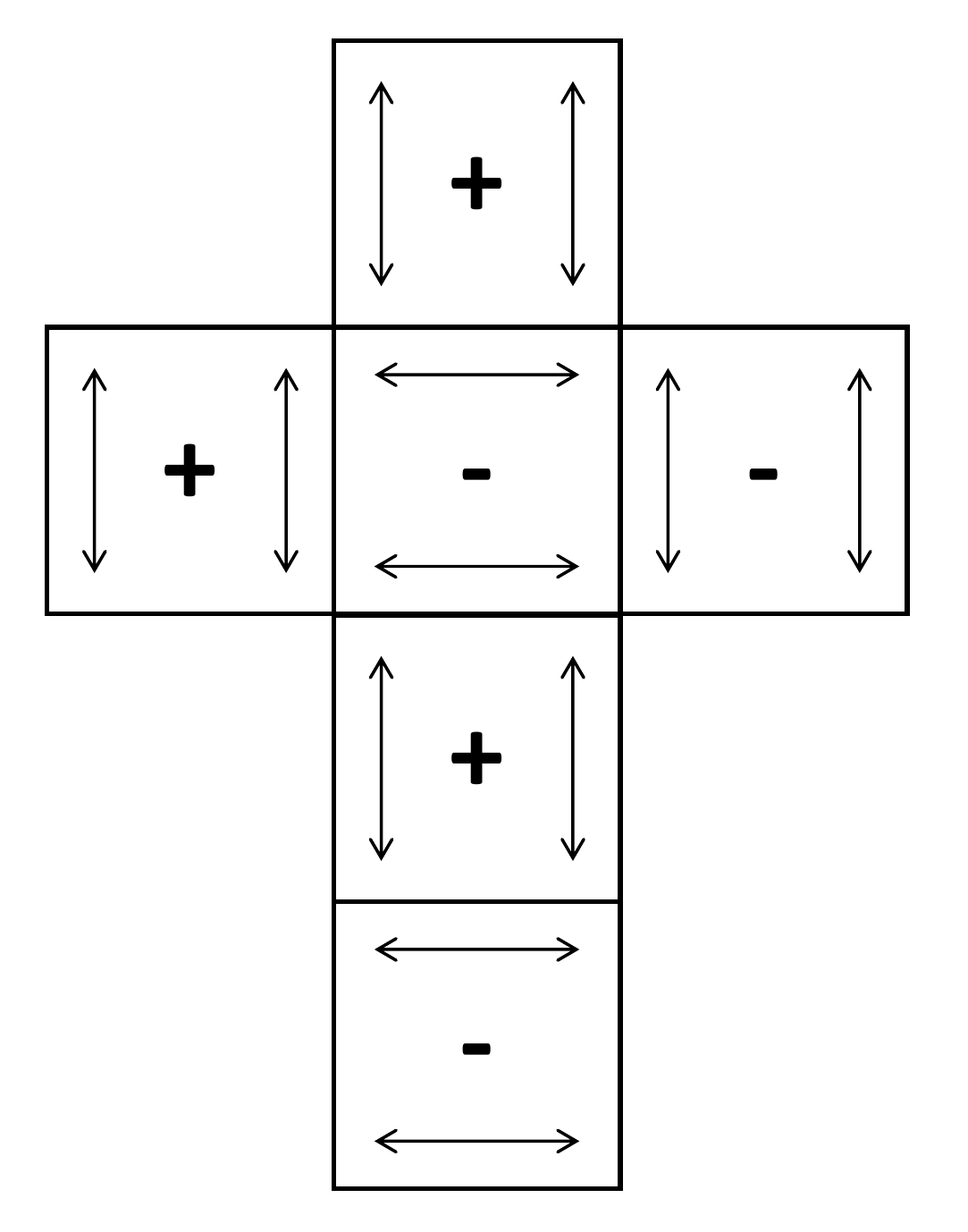}
\caption{The six-faces of the die which is used to create an entangled two-die system. The only difference with the die previously described in Fig.~\ref{Quantum-die-faces}, is in the permutation of two of the ``$+$'' and ``$-$'' symbols, whereas the relative orientation of the faces' surfaces has remained the same.
\label{Quantum-die-faces-bis}}
\end{figure}

To do so, we need to slightly modify the die we have previously defined in Fig.~\ref{Quantum-die-faces}. The only change we need to consider is the permutation of the ``$+$'' and ``$-$'' symbols on two of its faces (with no change of the corresponding orientations), so as to obtain the die described in Fig.~\ref{Quantum-die-faces-bis}.

\begin{figure}[!ht]
\centering
\includegraphics[scale =.6]{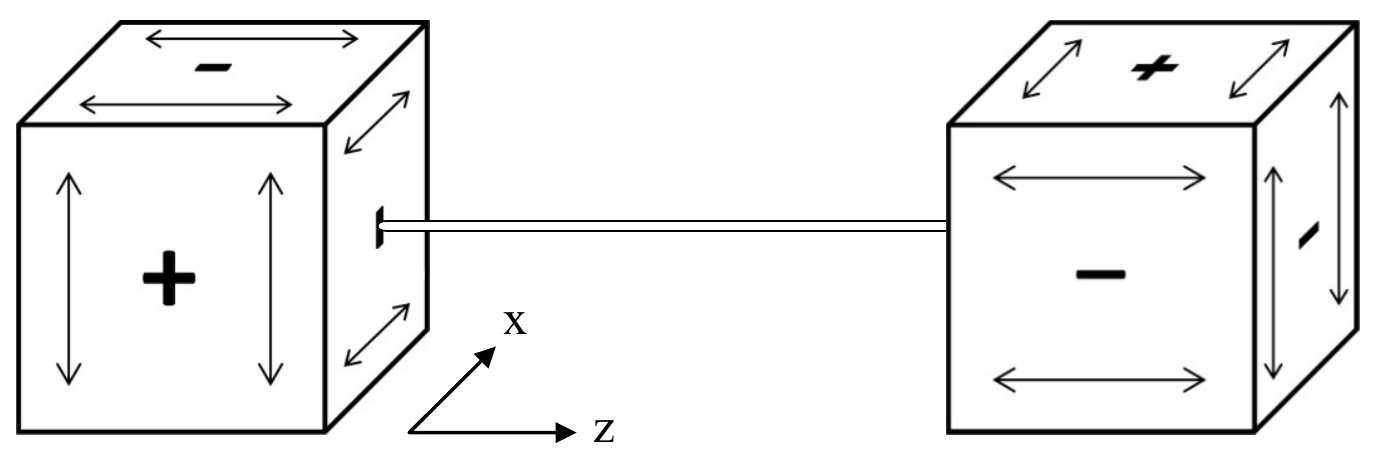}
\caption{The system of two (entangled) dice, connected through space by a rigid rod glued on two of their opposing faces.
\label{dice-connected}}
\end{figure}

Considering two identical dice of this kind, we can easily create an entangled double-die system by \emph{connecting them through space} by means of a rigid rod, whose two ends are glued at the center of two of the opposed faces of the two dice, as indicated in Fig.~ \ref{dice-connected}. 

\begin{figure}[!ht]
\centering
\includegraphics[scale =.5]{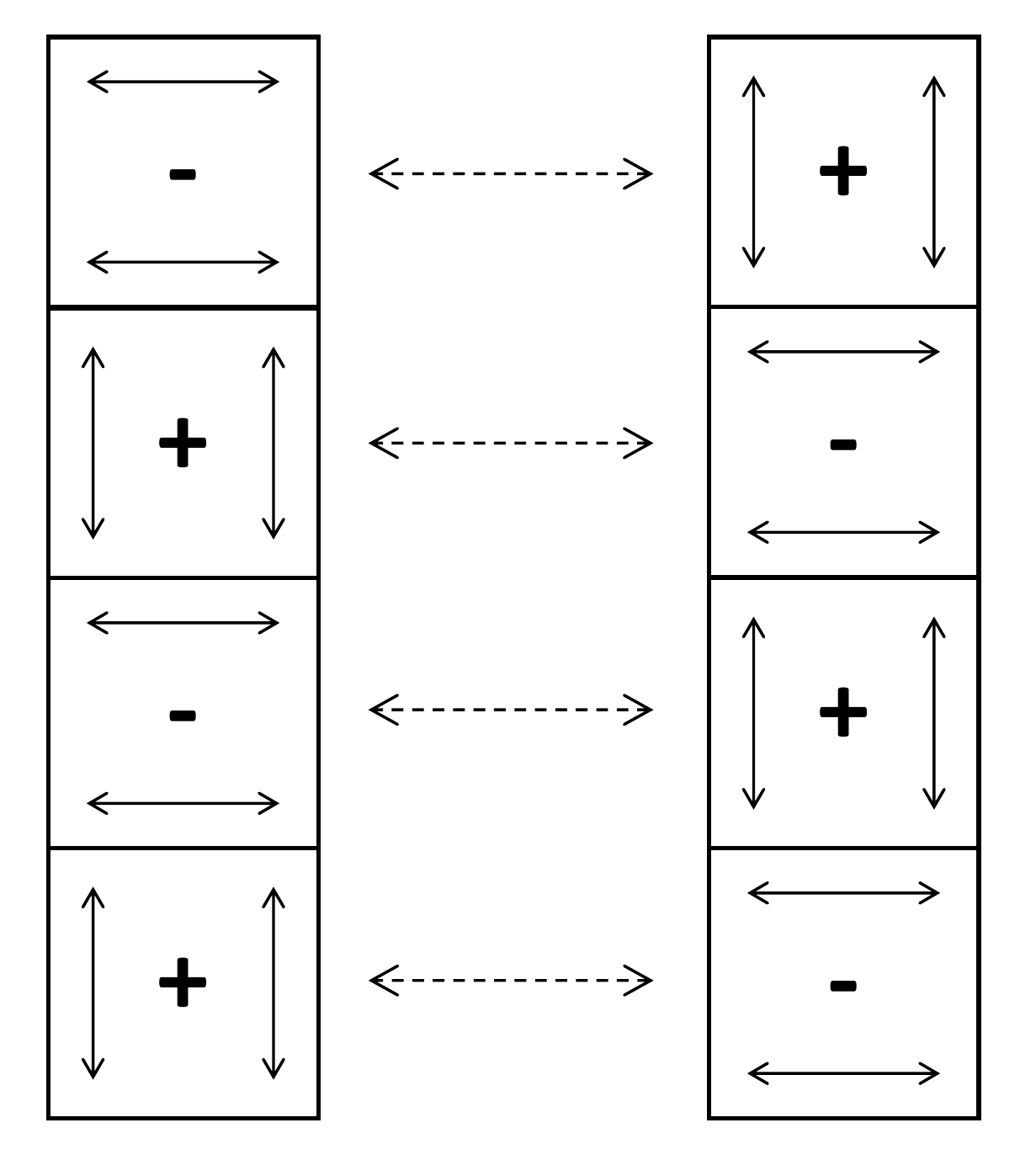}
\caption{The correlations between the four faces of the two dice whose normal vectors are orthogonal to the $z$-direction, due to the presence of the connecting rod.
\label{Dice-faces-correlations}}
\end{figure}

Clearly, the presence of the rod creates \emph{actual} correlations between the six different faces of the two dice. Here we are only interested in the correlations between the four faces of each die which correspond to a possible outcome (as a final upper face) in relation to a $x$-roll experiment. As it can be deduced from Fig.~\ref{Quantum-die-faces-bis} and Fig.~\ref{dice-connected}, the correlations in question are those described in Fig.~\ref{Dice-faces-correlations}.

The length of the rod is of course arbitrary. We only assume that it is made of a material which is sufficiently light in comparison to the mass of the two dice, and extremely rigid as well. In fact, the rod is not an essential ingredient in our analysis: we just use it to help us visualize the two dice as two spatially separated entities, and clearly identify in the rod the source of their connection through space. But we could very well avoid the use of the rod by for instance directly gluing together the two opposing faces of the two dice.

Having said that, we assume that the glue used to connect the rod is sufficiently strong, so that if the two dice are rolled together, simultaneously, in the same $x$-direction, they will be able to maintain their connection while rolling, i.e., to remain a whole entity. But we also assume that the glue, although strong, is not as strong as to allow the two dice to remain connected if only one die is rolled at a time (if the two dice are, say, made of metal, then instead of the glue we can imagine using a rod of a magnetic kind).   

In other terms, if we apply the shooter to only one of the two dice, to produce a $x$-roll, then, because of the inertia of the other die, the impact will cause the rod to suddenly detach and fall, thus disconnecting the two dice (one of which will be rolling, or sliding, whereas the other one will remain essentially still). On the other hand, if two shooters are used at the same time, on both dice, the torque experienced by the rod will be much lower, so that it will not detach and the two dice will be able to roll together on the game table, as a one piece entity (the double-die cannot slide, but only roll, as one of its two lower faces is always a high-friction face).

Keeping in mind the above, we now assume that two players are placed each one close to one of the two dice, who we shall call player $A$ and player $B$. Player $A$ performs on its dice (say, the left one) experiments $e^A_a$ and $e^A_{a'}$, which are defined as follow. 

Experiment $e^A_a$ consists in observing $F_x$, i.e., in using a shooter to produce a roll of the die along the $x$-direction, then reading the number marked on the obtained die's upper face, producing in this way one of the two outcomes: $o^A_a=+1$, or $o^A_a=-1$. Experiment $e^A_{a'}$ is much simpler, as it consists in simply looking at the die's upper face and checking whether it is flat or not. If it is so, then the outcome is $o^A_{a'}=+1$, otherwise it is $o^A_{a'}=-1$. 

Player $B$ performs on its dice (the right one) the same experiments as player $A$. In other terms, $e^B_b$ is defined as $e^A_a$, and $e^B_{b'}$ as $e^A_{a'}$. 

Of course, since all the faces of the two dice are by definition flat, and that the only faces of the two dice which are oriented toward the $x$-direction are those with a ``+'' symbol, all of the four above mentioned experiments, when singly performed, can only produce the outcome $+1$.  The same remains obviously true when the coincidence experiments $e_{ab'}^{AB}$, $e_{a'b}^{AB}$ and $e_{a'b'}^{AB}$ are performed at the same time by the two players, whose outcomes are always $(+1,+1)$.   

The situation changes however when one considers the coincidence experiment $e_{ab}^{AB}$, which \emph{creates upper faces' correlations}. Indeed, if the two players use simultaneously a shooter to impart a $x$-roll to their respective dice, then, as we explained, the rod will not separate and the two dice will remain connected as they roll. Therefore, according to Fig.~\ref{Dice-faces-correlations}, the only possible outcomes of the coincidence experiment $e_{ab}^{AB}$ are $(+1,-1)$ and $(-1,+1)$, and of course they have the same probability to occur, which is equal to $1/2$.

According to (\ref{expectation value}), we thus obtain that $E^{AB}_{ab'}= E^{AB}_{a'b}=E^{AB}_{a'b'}=1$, and $E^{AB}_{ab}=-1$, so that:
\begin{equation}
\label{Bell violation classical}
|E^{AB}_{ab} - E^{AB}_{ab'}| + |E^{AB}_{a'b'} + E^{AB}_{a'b}| = \left|-1-\left(+1\right)\right|+\left|+1+\left(+1\right)\right| =4.
\end{equation}

In other terms, not only the double-die system breaks Bell's inequality, but it does so in a maximal way. Before discussing the physical content of the above violation, let us just emphasize that the fact that the violation is maximal is only due to the fact that the connection between the two dice is such that the outcomes $(+1,+1)$ and $(-1,-1)$ are impossible, and consequently the difference $|E^{AB}_{ab} - E^{AB}_{ab'}| = |E^{AB}_{ab} - 1|$ in (\ref{Bell violation classical}) necessarily takes its maximal value. However, it is possible to use two dice of a more general geometry, like two prisms with an arbitrary number of faces, to produce weaker violations of inequality (\ref{Bell inequalities})~\cite{Massimiliano-preprint}.

\subsection{Discussion}
\label{Discussion2}

Let us now discuss the physical content of our experiment with the double-die system, to see what it reveals as regards a possible mechanism responsible for the violation of Bell's inequality. But before doing so, we would like to mention that it was Diederik Aerts who, many years ago, challenged for the first time the widespread belief that quantum structures would be only present at the microscopic level of our realty. He did that not only by showing that one can conceive ``classical'' machines exhibiting the typical (non-Kolmogorovian) probabilistic structure possessed by microscopic systems~\cite{Aerts3, Aerts4, Aerts4b}, but also, as we have done in the second part of this article, that one can use ordinary macroscopic entities to violate Bell's inequality (here the CHSH version of it).

The historical model used by Aerts to violate Bell's inequality was a machine made by vessels, tubes and water, known as the \emph{connected vessels of water} model~\cite{Aerts5, Aerts2} (an alternative, simplified version of such model, using a single uniform elastic band, has also been recently described by this author.~\cite{Massimiliano-elastic}). These models, like our two-die system, violate the inequality in a maximal way ($I=4$). However, Aerts was also able to conceive more elaborated macroscopic systems which can violate the inequality exactly in the same way as is done by a photon in a singlet state ($I=2\sqrt{2}$).~\cite{Aerts-rod, AertsBroekaert} 

Having said that, let us now analyze what our model teaches us regarding the nature of the correlations involved in the violation of Bell's inequality. Here we must distinguish between two different sorts of correlations: \emph{correlations of the first kind}, which are already present in the system before the execution of the experiment, and \emph{correlations of the second kind}, which aren't present before the execution of the experiment, but are literally created by it.~\cite{Aerts2} 

As far as this author can judge, a quite widespread belief is that Bell's inequalities would be violated because of the presence in the entangled state of correlations of the first kind, which therefore are only discovered (and not created) by the coincidence measurements. This belief appears to be supported by the observation that in Aspect's famous polarization experiments with entangled photons in singlet states,~\cite{Aspect1, Aspect2} it was possible to take the precaution to change randomly, in very short times, the orientations of the polarizers during the flight of the two entangled photons, thus enforcing relativistic separation between them. This means that, according to relativity theory, any form of interaction/communication \emph{through space} between the two measured subsystems has been excluded, and therefore it is quite natural to conclude that the observed correlations can only be of the first kind. 

This conclusion appears however to be in contradiction with the observation that a singlet state is a rotational invariant state, so that it  doesn't describe the state of two-entities having already actualized their respective polarizations, although the way it is mathematically written may wrongly suggest so.

So, what does our double-die system tells us about this subtle distinction between correlations of the first and second kind, and their role in the violation of Bell's inequality? At first sight the model seems to confirm the belief that correlations of the first kind would be responsible for the violation. Indeed, because of the presence of the rod, the faces of the two dice are clearly all correlated, and such correlations clearly exist even before the coincidence experiment $e_{ab}^{AB}$ is executed. But we must here properly distinguish ``faces'' from ``upper faces.''   

As we discussed at some length in Sec.~\ref{Rolling a die is a quantum process}, a rolling experiment involves an unmistakable creation aspect: the creation of a specific \emph{upper face} from the many available faces of the die, which are all obviously existing prior to the experiment as ``faces,'' but certainly not as ``upper faces.'' So, in the same way that by rolling a die we create an upper face, which did not exist prior to the rolling experiment, when we roll a system made of two connected dice we create a correlation between two upper faces, which did not exist prior to the experiment. 

The subtle point here is that during a coincidence experiment the correlation between two ``upper faces'' is created from the existing correlations between ``faces,'' i.e., from the correlations between ``\emph{potential} upper faces,'' which therefore can also to be understood as \emph{potential correlations} between ``upper faces.'' In other terms, if correctly interpreted, our double-die model demonstrates that it is indeed the mechanism of \emph{creation of correlations} (correlations of the second kind, as Aerts proposed to call them~\cite{Aerts5, Aerts2}) which is responsible for the violation of Bell's inequality. 

To convince oneself that this is indeed the correct interpretation, one can try to use only correlations of the first kind to violate Bell's inequality in our double-die model, and observe that it is impossible. One can do that by simply replacing the observable $F_x$, which is responsible for the creation of an upper face (through the $x$-roll), by an observable $\tilde F$, of a purely discovery kind, consisting in simply taking notice of the already actual upper face showed by the die. One can easily check then, that if $\tilde F$ is used instead of $F_x$, then, depending on the state in which the system is prepared, one obtains either $I=0$ or $I=2$, in accordance with Bell's inequality. 

So, our model confirms what has been repeated several times by Aerts~\cite{Aerts-rod}: ``The possibility of violating Bell inequalities is not only a property of quantum entities. Bell inequalities can also be violated by coincidence measurements on a classical macroscopical entity. In fact Bell inequalities can always be violated if during the coincidence experiments one breaks one entity into separated pieces, and by this act creates the correlations.''

However, differently from the macroscopic models that have been studied in the past, in our model it is not the fact that the entity is broken that creates correlations, but, on the contrary, the fact that it is not broken!

\section{Concluding remarks}
\label{Concluding remarks}

In this paper we have didactically introduced the reader to some important ideas regarding the possibility of a realistic interpretation of the behavior of microscopic quantum systems. We have done so by somehow reverting the logic of Einstein's celebrated quote, that God doesn't play dice, showing that the simple act of rolling a die (according to certain protocols) is a truly quantum experiment, which can be described using the projection postulate and the Born rule, and which is capable to produce interference effects.  

This allowed us to gain some intuition into a possible origin of quantum probabilities, which can be understood as epistemic statements associated to our lack of knowledge not about the state of the system, but about the exact interaction taking place between the system and the measurement apparatus, according to Aerts' hidden-measurement approach.  

This possibility, of understanding quantum probabilities not as irreducible (ontic) quantities, but as contextual (epistemic) quantities, allows us to demystify much of the mystery associated to the quantum measurement, which can be understood as a physical (and not psychophysical) creation process induced by the interaction of the system with the measuring apparatus. Also, it shows that quantum structures are not limited to the microworld, but are also present in macroscopic systems, if we only limit in a certain way our possibilities of actively experimenting on them, according to specific experimental protocols. 

Another interesting aspect indirectly touched by our analysis is the possibility of generally understanding probability theory as a theory dealing with the measurement of properties associated to physical systems, operationally defined by means of certain specific observational protocols.  If we understand probabilities in this way, many interpretational difficulties immediately disappear. For instance, in so-called Bertrand paradox,~\cite{Bertrand} the fact that different randomization procedures yield different probabilities can simply be understood as the measurement of different properties (or observables), associated to different observational protocols, and therefore to different sets of hidden-measurement interactions.

In that respect, it is worth observing that the hidden-measurement mechanism is in fact a very general one, in the sense that it can be used to describe any probabilistic situation, and not only the typical quantum ones. In other words, it is able to provide, in a way, a full description of all type of probability structures one can encounter in the world, as established in Ref.~\cite{Aerts-1994} (Sect. 4).

In the second part of this article we have studied another important and mysterious feature of quantum systems: entanglement. We have done so by showing that Bell's inequality can easily be violated by macroscopic systems, provided the measurement process can create correlations, and not just discover correlations. Our double-die system cannot however be described by using a Hilbert space formalism and self-adjoint observables, as is clear that $I=2\sqrt{2}$ is the maximal possible violation in this ambit~\cite{Cirel}. 

On that respect, it is worth emphasizing that, according to a theorem of Pitowski ~\cite{Pitowski}, when  Bell-type inequalities are violated, then one cannot use a classical Kolmogorovian probability model to represent the probabilities associated with the experiments under consideration. Quantum microscopic systems are of course important examples of non-Kolmogorovian probability models, but they are not the only ones, as the example of the double-die, violating Bell's inequality in a maximal way, clearly shows. 

That said, let us conclude by observing that although the double-die system certainly elucidates a possible mechanism behind the violation of Bell's inequality, what it doesn't reveal is how quantum systems are able to implement such mechanism. Indeed, the reason for the creation of correlations in the double-die is of course the presence of the connecting rod: because of it we cannot consider the two dice as two spatially separated entities, but as a whole non separable entity. This \emph{macroscopic wholeness} property~\cite{Aerts2} of the double-die system is of course totally without mystery, being the result of their connection through space by means of the glued rod. 

On the other hand, although it is also possible to conclude with Aspect that~\cite{Aspect2} ``[...] an entangled EPR photon pair is a non-separable object; that is, it is impossible to assign individual local properties (local physical reality) to each photon,'' what is much more difficult to understand is how the two photons can actually remain connected, considering that they do not possess the property of macroscopic wholeness.

This apparent paradox, of a microscopic entity made of a pair of spatially separated entities which can nevertheless remain experimentally connected, independently of their spatial distance, is of course what fundamentally distinguishes our double-die macroscopic system from an entangled microscopic system. Another important distinction is of course the fact that for a rod to allow quasi-instantaneous correlations between the two dice (as it happens between two correlated pairs in Aspect's experiments), when these are separated by a very large spatial distance, it should become an arbitrarily rigid and light object (the typical mechanical properties attributed in the past to the luminiferous ether), which of course is impossible to achieve with a concrete macroscopic object. 

Now, it is precisely because of this conceptual difficulty, of having to give sense to a connection through space which cannot be detected nor imagined in a non-magical way, that the general belief among physicists is that correlations of the first kind would be responsible for the violation of Bell's inequality in experiments with entangled pairs, like those conducted by Aspect. However, this belief cannot be considered satisfactory, as we are not able to conceive -- as far as the present author knows -- macroscopic models that would be able to violate Bell's inequality by only using correlations of the first kind. This appears to be the case if the double-system in question obeys Bell's locality assumption, as is the case of the enigmatic macroscopic device described by Mermin~\cite{Merm}, but also if the locality is manifestly disobeyed, as our two-die system (and similar interconnected systems~\cite{Aerts5, Aerts2, Massimiliano-elastic}) clearly show.

This explains why we have so strongly emphasized in this paper the deep conceptual difference between a ``face'' and an ``upper face'' of a die. Indeed, if we don't properly understand such a fundamental distinction, we are easily led to believe that macroscopic systems like a double-die would actually demonstrate the possibility of violating Bell's inequality by means of correlations of the first kind, which on the contrary is not the case, as we have explained in the second part of our paper. 

Having said this, let us point out a possible solution of this apparent conundrum, which consists in simply abandoning the preconception that microscopic entities would always be present in our three-dimensional space, i.e., that physical reality should only exist within space. In other terms, it is about accepting that, quoting here Aerts, space would only be~\cite{Aerts4} ``[...] a momentaneous crystallization of a theatre for reality where the motions and interactions of the macroscopic material and energetic entities take place. But other entities -- like quantum entities for example -- `take place' outside space, or - and this would be another way of saying the same thing -- within a space that is not the three dimensional Euclidean space."

If we accept the idea that \emph{non-locality} is actually an expression of \emph{non-spatiality},~\cite{Aerts4, Aerts2, Massimiliano1, Massimiliano2, Massimiliano3, Massimiliano-God} then of course there are no conceptual problems in considering that two non-spatial microscopic entities could remain, as time passes by, intimately connected (not ``through space'' but, more generally, ``through reality''!), and that it would be their non-spatial connection the responsible for the \emph{creation of correlations} that violate Bell's inequality. Also, if the connection is assumed to be non-spatial, then the process of creation of correlations needs not be limited by relativistic constraints, as is the case for macroscopic objects, compatibly with the superluminal correlative effects observed in EPR-like experiments.

\end{document}